\documentclass{article}
\usepackage{spconf,amsmath,graphicx}
\usepackage{bm}

\title{singing voice conversion with non-parallel data}
\name{Xin Chen$^1$, Wei Chu$^1$, Jinxi Guo$^2$, Ning Xu$^1$}
\address{$^1$Snap Research, Snap Inc., USA\\$^2$University of California, Los Angeles, USA \\
\texttt{\normalsize \{xin.chen, wei.chu, ning.xu\}@snap.com, lennyguo@g.ucla.edu}}
\begin{document}
%
\maketitle
\begin{abstract}

Singing voice conversion is a task to convert a song sang by a source singer to the voice of a target singer. In this paper, we propose using a parallel data free, many-to-one voice conversion technique on singing voices. A phonetic posterior feature is first generated by decoding singing voices through a robust Automatic Speech Recognition Engine (ASR). Then, a trained Recurrent Neural Network (RNN) with a Deep Bidirectional Long Short Term Memory (DBLSTM) structure is used to model the mapping from person-independent content to the acoustic features of the target person. F0 and aperiodic are obtained through the original singing voice, and used with acoustic features to reconstruct the target singing voice through a vocoder. In the obtained singing voice, the targeted and sourced singers sound similar. To our knowledge, this is the first study that uses non parallel data to train a singing voice conversion system. Subjective evaluations demonstrate that the proposed method effectively converts singing voices. \end{abstract}

%

\section{Introduction}\label{sec:introduction}

Singing is widely employed in most cultures as means of entertainment and self-expression. It should also be noted that singing is an important way to convey linguistic information. Singing voice conversion, a task to convert a song sang by a source singer to the voice of a target singer, has many practical applications.  For instance, a user can first sing a song and then replace his voice with another person's voice. This potential use demonstrates a fun and creative way to generate unique, collaborative content. Moreover, the user can pretend that he or she is singing a song by replacing the original singer's voice with his or her own voice, displaying his or her singing on social media platforms. 

Singing voice conversion and conventional speech voice conversion are similar tasks. In general, these two types of voice conversions need to divide content into person-dependent and person-independent content. Both singing and speech voice conversions switch the person-dependent content from source to target and retain person-independent content. However, in speech voice conversion, the manner of speaking (including the speech pattern, pitch, dynamics, duration of words, etc.) contains important information about the speaker. Therefore, the manner of speaking belong to person-dependent content and need to be modeled and changed from the source speaker to the target speaker. On the other hand, in singing voice conversion, the manner of singing is primarily determined by the song itself. Consequently, the manner of singing should be considered person-independent content and remain a key part of the singing voice conversion process. Only characteristics of voice identity such as the timbre are considered person-dependent content and need to be replaced. 



Various singing voice conversion methods are proposed to convert the singing voice from one to another \cite{Kobayashi2015} \cite{SingingVoice2014} \cite{SingingVoice2010}. Parallel data is generally required to model the singing voice conversion. While voice conversion has recently gained popularity, typical voice conversion training requires parallel data \cite{Voiceconversion1998} \cite{Voiceconversion2007} \cite{Voiceconversion2013} \cite{Voiceconversiondeepnets2013}. Since it is often difficult to obtain parallel data for speech voice, various techniques for non-parallel training voice conversion \cite{Voiceconversion2010} have been proposed. Experimental results show that the performance of these latter techniques is inferior to those of the VCs with parallel data. This outcome may be explained by the difficulty to accurately perform alignment on non-parallel data. More recently \cite{aryal2015articulatory} a newly-proposed approach mapped electromagnetic articulography (EMA) features to speech for a foreign accent conversion task. In \cite{Liu2015}, the authors proposed using Phoneme State Posterior Probabilities (PSPP) for speaker independent content modeling and lip motion animation. In \cite{Sun2016}, the authors proposed model content use of Phonetic Posterior Grams (PPG) to encode the speaker independent content and to map this feature to speech for voice conversion. 

In this paper, we propose using a parallel data free, many-to-one technique and using phonetic posteriors as the major person-independent content for singing voice conversion. To our knowledge, this is the first study that uses non parallel data to train singing voice conversion models. In the training stage, we only use some unlabeled target speech data that is relatively easy to obtain. We first decode the speech data into a phonetic posterior probability sequence using a robust Automatic Speech Recognition Engine (ASR). These phonetic posterior probability sequences contain only the content of the speech data and no user identity information. From the speech data we also apply parameter analysis to extract acoustic features. Parameter analysis contains both speech content and the speaker characteristics in order to reconstruct the speech via a vocoder. Those phonetic posteriors and acoustic features are used as input and output to train a Recurrent Neural Network (RNN) with a Deep Bidirectional Long Short Term Memory (DBLSTM) structure. As a result, this process builds a mapping from the person-independent phonetic posteriors to acoustic features that contain both person-dependent and person-independent content. In the conversion stage 1, a phonetic posterior sequence that encodes the person-independent content is generated by decoding a singing voice through the ASR. In stage 2, the trained DBLSTM-RNN is used to map the phonetic posterior to the acoustic features of the target singing voice. F0 and aperiodic are obtained through the original singing voice, and used together with acoustic features to reconstruct the target singing voice through a vocoder. 


The paper is organized according to the following sections. In Section 2, we describe the Deep Neural Network(DNN) model for singing voice recognition.  In section 3, we describe the DBLSTM-RNN structure used to model the map between encoded phonetic posteriors and acoustic features. Section 4 offers an in-depth description of the training and conversion stages of the proposed method. We then describe our experiment set up and demonstrate subjective evaluation results in Section 5, and conclude our work in Section 6. We also include a selection of samples.\footnote{https://sites.google.com/site/singingvoiceconversion2018/}

\section{Deep Neural Network acoustic model for singing content recognition}

In \cite{Gruhne2007}, the authors proposed using various classifiers such as a Support Vector Machine (SVM), Multi Layer Perceptron (MLP) and Gaussian Mixture Model (GMM) for phoneme recognition on a singing voice, observing that the recognition accuracy improved with harmonics analysis. In \cite{Mesaros2012} \cite{Mesaros2009}, the authors proposed the use of a traditional HMM/GMM acoustic model with MLLR adaptation to enable automatic recognition of lyrics expressed in a singing voice. Recently, Deep Neural Network (DNN) based acoustic modeling has demonstrated superior performance when compared to the traditional method \cite{hinton2012deep} and has become the state of the art technology in speech recognition. It is therefore of interest to apply this technology to transcribe singing voice into a phonetic posterior probability sequence in order to encode the content.

As a brief review, a DNN is a multi-layer perceptron (MLP) with many hidden layers. Each hidden layer computes the activations of conditionally independent units given the activations of the previous layer. If we denote the input vector of a hidden layer $l$ as $\bm{x}_{l-1}$, then the output vector of the layer $l$ can be computed as 
\begin{equation}
\label{eq.nn}
\bm{x}_l = \sigma (\bm{W}_l \bm{x}_{l-1} + \bm{b}_l),
\end{equation}
$\bm{W}_l$ and $\bm{b}_l$ are the weight matrix and bias vector of layer $l$, and $\sigma$ is the predefined activation function. Choosing $\sigma$ as nonlinear functions may allow networks to model nontrivial problems. There are multiple layers in order to model complex signals such as speech.

To model the probabilities for the phoneme class vector $\bm{s}$, the softmax activation function is predominantly used in the last layer of a DNN: 
\begin{equation}
\label{eq.softmax}
P(\bm{s}|\bm{x_0}) = \frac{\exp(\bm{W}_n \bm{x}_{n-1} + \bm{b}_n)}{\sum_{s} \exp(\bm{W}_n \bm{x}_{n-1} + \bm{b}_n)}. 
\end{equation}
Since the vector $P(\bm{s}|\bm{x_0})$ sums to one and all its elements are between zero and one, $P(\bm{s}|\bm{x_0})$ represents a categorical probability distribution. 

\section{Deep Bidirectional LSTM - Recurrent Neural Network (RNN) }

BLSTM is a type of Bi-directional RNN that was firstly proposed in \cite{graves2005framewise}, which can make the best use of context in both directions. LSTM \cite{Schmidhuber1997} was used as the memory block. Inspired by the recent success of Deep learning models, a multiple stacked layer DBLSTM was introduced and yielded good performance in speech recognition \cite{Graves2013} and voice conversion \cite{Sun2016}. We first offer a brief review of this approach: 

\begin{figure}[h!]
 \centerline{\framebox{
 \includegraphics[scale=0.35]{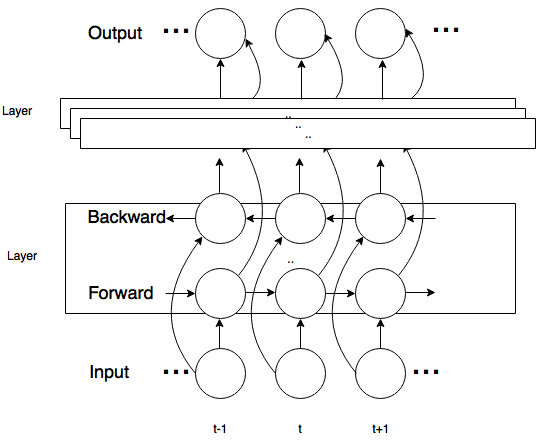}}}
 \caption{An illustration of a deep bidirectional recurrent neural network to map input sequences to output sequences.}
 \label{fig:bdlstm}
\end{figure}

\begin{figure}[h!]
 \centerline{\framebox{
 \includegraphics[scale=0.32]{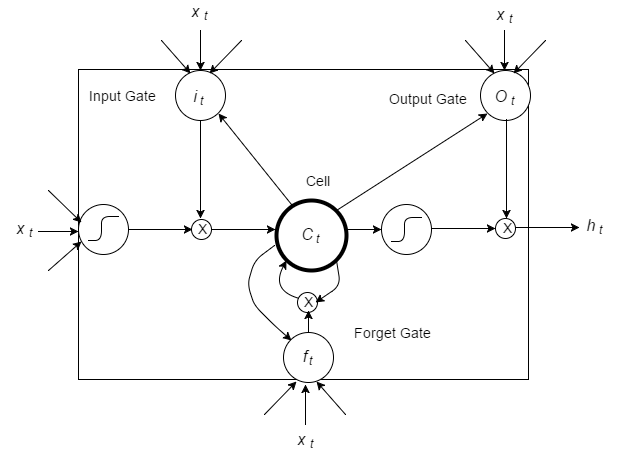}}}
 \caption{An illustration of a single Long Short Term Memory (LSTM) cell. }
 \label{fig:memoryblock}
\end{figure}

Given an input sequence, a proposed corresponding recurrent neural network calculates the hidden vector sequence and output sequence by iterating layer by layer, as illustrated in figure \ref{fig:bdlstm}. Each layer contains both forward feeding and backward feeding. Generally, multiple layers are stacked between input and output layers. Each cell represents a memory block, which is a Long Short Term Memory (LSTM) block. An illustration of LSTM block is presented in figure \ref{fig:memoryblock}. This architecture uses purpose-built memory cells to store information and exploit long range context, and is very powerful in presenting the mapping power between encoded phoneme sequences and corresponding acoustic features. 

\section{Singing Voice Conversion}
In this section we describe the technical steps of our singing voice conversion method. Since our method is parallel data free, we could use any speech data from the target to train a model in the training stage. This data is fully independent of the singing voice to be converted. We call this process the "many to one method" because it can be applied to any source singing voice after the model is trained. 

\subsection{Training stage}

\begin{figure}
 \centerline{\framebox{
 \includegraphics[scale=0.65]{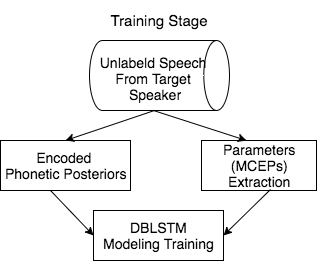}}}
 \caption{An illustration of the training stage. A BLSTM model is trained given phonetic posteriors and MCEPs generated from target speaker's voice data.}
 \label{fig:training_stage}
\end{figure}

In the training stage, we try to train a DBLSTM model to map the encoded phonetic posteriors to target speaker's acoustic features. For different target singers we need to train different models. As showed in figure \ref{fig:training_stage}, we only use unlabeled target voice data for training. For each utterance, first, a speaker independent DNN phoneme acoustic model is used to extract the posterior probabilities for each phoneme at each frame, and the information is encoded as a matrix to represent the content information of the giving segment of voice data. Second, an acoustic feature parameter extraction tool is used to extract Mel Cepstral (MCEP) feature. We collect a dataset of paired encoded content information and its corresponding acoustic features and use them to train the mapping with DBLSTM.   

\subsection{Conversion stage}

\begin{figure}[h!]
 \centerline{\framebox{
 \includegraphics[scale=0.6]{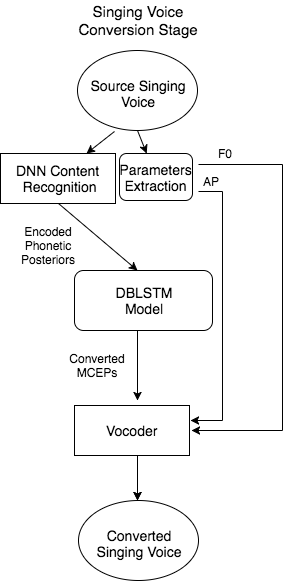}}}
 \caption{An illustration of the conversion stage. The trained DBLSTM model for target singer is applied to convert the phonetic posteriors extracted from the source singing voice to target's MCEPs for synthesizing target singer's sining voice.}
 \label{fig:testingstage}
\end{figure}

With a trained DBLSTM model, we can map the person-independent content to target singer's acoustic features and use that to synthesize new singing voice. As shown in figure \ref{fig:testingstage}, given a specific singing voice clip, a robust automatic speech recognition (ASR) engine is used to generate the encoded phonetic posteriors that contains singer-independent content of the singing voice clip. This encoded phonetic posteriors sequence is then mapped to the corresponding MCEP acoustic features of target singer using our DBLSTM model trained in the training stage for the target singer. From the source singing voice clip, F0 and Aperiodic information is also extracted by using a parameter extraction tool, and is kept unchanged. Taking these three piece of information, a vocoder is used to synthesize a singing voice that shares the voice identity of the target speaker while retaining the lyrics and melody espoused by the original singer.

\section{Experiments}
In this section we describe the technical details of how we conduct experiments and also the subjective evaluation of our results.

\subsection{Automatic Speech Recognition (ASR) module}
In our experiments, TIMIT database is used for training the phoneme recognition module using DNN. The training data set consists of 3696 utterances from 462 speakers. There are 39 phonemes. The acoustic features are 13 MFCCs that are extracted with a 5ms shift. A context of 17 Frames of feature, shaped by mean variance normalization, is used by DNN as the input. The input dimension is 221. The ASR DNN contains four fully connected layers, and each layer contains 2048 hidden nodes.  The output layer of the DNN contains 39 phoneme classes. TNet \cite{neutralnetwork2010} is used to conduct the training. The frame accuracy on validation set reaches 70.7\%.

\subsection{DBLSTM module}
In our proposed method, we will train a DBLSTM model for each target voice. Here we use CMU Arctic data for training our models for target voice, and a female voice SLT is used for the experiments in this paper. There is a total of 1132 utterances. Only speech data from the corpus is used. We choose a Mel Cepstrual feature (MCep) as the acoustic feature to be modeled within the DBLSTM structure, and set the dimension to 40. The feature is extracted by using World \cite{morise2016world}. The input dimension for the DBLSTM is 39, and represents 39 phonemes. We train two models: one model is 128N, trained with 400 Utterances, containing a stacked 4 layers with 128 hidden nodes in each layer; the other model is 512N, trained with all available utterances, containing a stacked 4 layers with 512 hidden nodes in each layer. The network is trained by using CURRENNT \cite{Opensource2015}.


\subsection{Subjective evaluation}

Our subjective evaluation is performed by 12 people listening on the original singing content and converted singing content, together with the target speakers voice. Those subjects are not professional in singing voice conversions. 

Several different types of source singing voices are used, some samples are randomly picked from the MIR 1K dataset \cite{hsu2010improvement}, including S01 (female) and S04 (male), who are both singing in Mandarin. S02 (male) is singing RAP in English, and S03 (male) is a general singing voice in English. We evaluated MOS on naturalness, score similarity (as showed in Table. \ref{tab:t2}). The scale is set between 1 to 5. 

\begin{table}[h!]
    \centering
    \begin{tabular}{c|c}
        \hline 
         Score & Meaning \\
        \hline
        1 & same to the original \\
        2 & similar to the original \\
        3 & in between \\
        4 & similar to the target \\
        5 & same to the target \\
         \hline
    \end{tabular}
    \caption{
    Rubrics for similarity
    }
    \label{tab:t2}
\end{table}

\begin{figure}[h!]
 \centerline{\framebox{
 \includegraphics[scale=0.38]{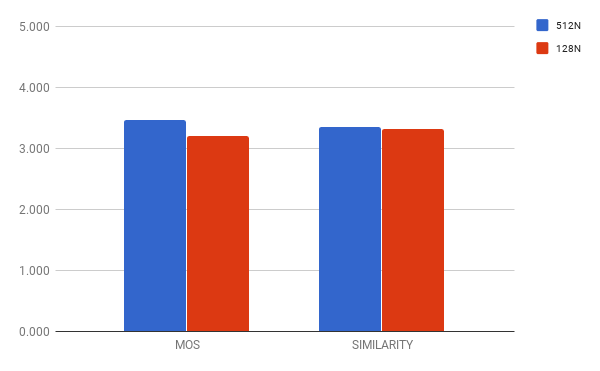}}}
 \caption{Subjective test results. Blue bars represent scores for 512N and red bars represent scores for 128N. }
 \label{fig:overall}
\end{figure}

In figure \ref{fig:overall}, the average MOS score of 128N model is 3.2, compared to that of the 512N model which scores a 3.4. This shows that a complex model with more training data generates slightly better naturalness. Additionally, our average similarity score of 128N model is 3.3 versus that of the 512N model which scores 3.4, showing our proposed approach is able to alter the voice identity towards the target speaker. Overall, the 512N model outperforms the 128N model, showing that more data and more complex model can help improve the performance.

\begin{figure}[h!]
 \centerline{\framebox{
 \includegraphics[scale=0.38]{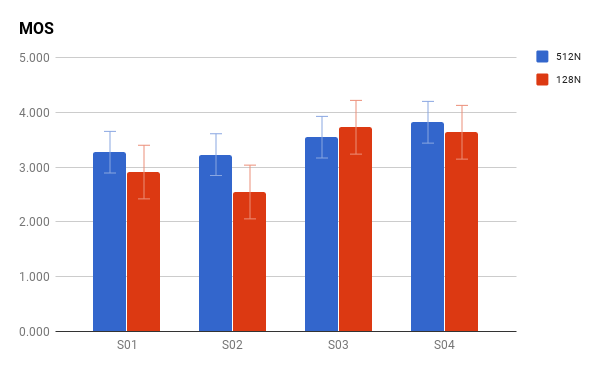}}}
 \caption{Subjective test results of individual MOS. Blue bars represent scores for 512N and red bars represent scores for 128N. Error range is shown on top of the bars. }
 \label{fig:mos}
\end{figure}

\begin{figure}[h!]
 \centerline{\framebox{
 \includegraphics[scale=0.38]{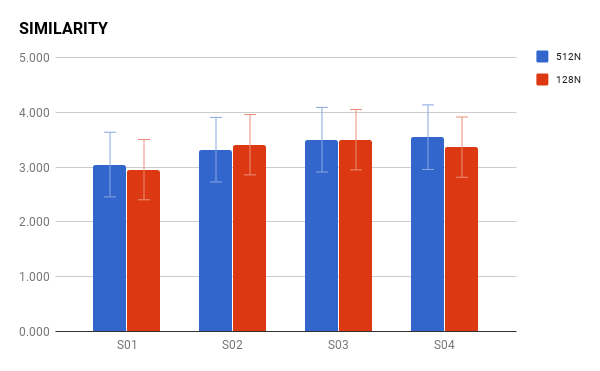}}}
 \caption{Subjective test results of individual similarity. Blue bars represent scores for 512N and red bars represent scores for 128N. Error range is shown on top of the bars.}
 \label{fig:similarity}
\end{figure}

Figure \ref{fig:mos} shows MOS score of each singing voice samples for both 128N and 512N models. Figure \ref{fig:similarity} illustrates the similarity score of each singing voices for both the 128N and 512N models. 

We found RAP audio clip (S02) achieves the lowest MOS score,  and the explanation could be the rapid change of speech content due to fast speaking rate dragging down the ASR performance. We also found male singer in our experiment tend to have higher similarity score, this could be due to the bigger difference between the source and target speaker, which is a female voice.

\section{Conclusion}

We propose a novel system to use a parallel data free, many-to-one voice conversion on singing voice conversion. A speaker independent ASR is first used to extract the phonetic posteriors sequence to represent the person-independent content, and a DBLSTM model is used to model the mapping from the person-independent content to target speaker's acoustic features. These acoustic features are used to synthesize the target singing voice via a vocoder, together with the F0 and aperiodic information extracted from the source content. 

To our knowledge, this is the first attempt to use non-parallel data to train a model for singing voice conversion. Additionally, subjective evaluation reveals that the proposed method is effective without using parallel data. 

For future enhancement, the authors would like to collect more singing voice data for model adaptation in speech recognition to further improve performance. In our future work, we also hope to explore neural vocoders such as wavenet \cite{VandenOord2015}, a recently proposed method that demonstrates superior performance when compared to traditional vocoders. 
\vspace*{20mm}

\bibliographystyle{IEEEbib}

\end{document}